\documentstyle[12pt,a41,epsfig]{article}

\begin{document}
\sloppy
\thispagestyle{empty}

\mbox{}
\vspace*{\fill}
\begin{center}
{\LARGE\bf HEMAS: a Monte Carlo code for hadronic,} \\

\vspace{2mm}
{\LARGE\bf electromagnetic and TeV muon components in air shower}\\

\vspace{2em}
\large
Eugenio Scapparone
\\
\vspace{2em}
{\it  Laboratori Nazionali del Gran Sasso,}
 \\
{\it S.S. 17 km 18+910 61070 Assergi(AQ), 
Italy}\\
\end{center}
\vspace*{\fill}
\begin{abstract}
\noindent

The features of the HEMAS code are presented. The results of the comparison 
between the Monte Carlo expectation and the experimental data are shown.
\end{abstract}
\vskip 5 cm
\centerline{ Invited Talk at ''International Workshop on Simulation and}
\centerline{  analysis methods for large neutrino detectors'',}
\centerline{ Desy Zeuthen, Berlino ( Germany), July 6-9, 1998}
\centerline{ To be published in the Proc. of the Workshop}

\vspace*{\fill}
\newpage
%
\section{Introduction}
Cosmic ray physics at energy E$\geq$10-100 TeV, due to the steepening of the
spectrum, can be performed 
only by using indirect measurements. An impressive amount of data has 
been collected by extensive air shower arrays, Cherenkov
detectors and underground muon experiments. The challenge is the interpretation
of these results. The bulk of the analysis are performed by assuming a given
Cosmic Ray spectrum and chemical composition (trial model), simulating the particle 
interaction and the shower development in atmosphere and finally comparing the 
simulated results with the real data. 
The reliability of the Monte Carlo simulation used is therefore a primary 
task for the
correct interpretation of these data: such difficulty stimulated a lot of 
experimental work to validate the existing model and many
theoretical ideas to improve the simulation tools. 
Modelling a Monte Carlo to describe the high energy cosmic ray interactions in
atmosphere is a hard task, since 
Cosmic Rays studied with indirect
measurements extend to energy and kinematical regions non covered
by the accelerator experiments yet.
Morever nucleus-nucleus collision have been investigated just up to 
few hundreds of GeV/nucleons. This poorness of experimental data is 
reinforced by 
the lack of a completely computable theory for the bulk of hadronic 
interactions, since QCD can be used only for high $p_{t}$ phenomena.

Many models have been developed in the last years, with different enphasis
on the various components of the C.R. induced shower.
Basically they can be splitted in two categories: the models using
parametrization of collider results(NIM85,HEMAS) and the phenomenological 
models inspired for istance to the Dual Parton Model or similar
approaches(DPMJET,SYBILL,QGSJET).
I will concentrate in this talk on the HEMAS code\cite{hemas}, stressing the
results of the comparison with the experimental data. 
\section{HEMAS: description of the code}
The HEMAS code was developed in the early '90, when a new generation 
of experiments (LVD, MACRO, EAS-TOP) were starting the data taking at Gran
Sasso. This code is suited to simulate high energy muons ($E_{\mu}$$\geq$500GeV) and
the electromagnetic size of the shower.
It is a phenomenological model, based on the parametrization of the 
collider data. The code describes multiple hadron production by means 
of the multicluster
model, suggested by UA5 experiment. 

The total p-Air cross section is one of the most important ingredients
of the codes. Since the cross section of hadrons on nuclei is not measured
directly at energies greater than several hundred of GeV, an
extrapolation to higher energies is required and is performed in the
context of "log(s)" physics. 

Figure 1 shows the
HEMAS cross section p-Air as a function of the centre of mass energy $\sqrt s$
compared with the cross section used in other Monte Carlo codes.
                                                                      
Given the $\sqrt s$ of the
interaction, the average number of charged hadrons $<$$n_{ch}$$>$ is choosen according with the equation:
\begin{equation}
<n_{ch}>=-7.0+7.2s^{0.127}.\\
\end {equation} 
The  actual number of charged hadrons $n_{ch}$ is sampled from a
a negative binomial distribution with
\begin{equation}
k^{-1}=-0.104+0.058ln(\sqrt s)
\end {equation} 
Respect to the previous codes, where 
$n_{ch}$ was sampled according to a poissonian, this choice reflects in a
larger  fluctuation of underground muon multiplicity.
Particles are then grouped in 
clusters, eventually decaying in mesons.

A relevant feature of HEMAS is the parametrization of muon parent mesons 
$p_{t}$ distribution. While for single pion cluster $p_{t}$ is always sampled 
from an exponential, for kaon clusters, for the leading particle and for 
pion clusters with at least two particles, $p_{t}$ has a given probability
to be extracted from a power low:\\
\begin{equation}
\frac{dN}{dP_{t}^{2}}=
\frac{const}{(p_{t}^{o}+ p_{t})^\alpha}
\end {equation}  
where $p_{t}^{0}$=3 GeV/c while $\alpha$ decreases logarithmically with 
energy 
\begin{equation}
\alpha=3+
\frac{1}{(0.01+0.011ln(s))}
\end {equation} 
Nuclear target effects are included too. The transverse momentum distribution 
is increased in p-N collision respect to the p-p case, according to the so
called 'Cronin effect'\cite{cronin}. 
The ratio R($p_t$) of the inclusive cross section on a target of mass A to 
that on a proton target depends in principle from the particle produced.
In HEMAS, R($p_{t}$) has been approximated with a single function:\\
\begin{equation}
R(p_{t})=(0.0363p_{t}+0.0570)K~~~~~for~~~ p_{t}\leq4.52 GeV/c
\end {equation} 
\begin{equation}
R(p_{t})=0.2211K~~~~~~~~~~~~~~~~~~~~~~ for~~~ p_{t}>4.52 GeV/c
\end {equation} 
where K is a normalization constant.

The average $<$$n_{ch}$$>$ in p-Air collisions is
obtained using the relation between the rapidity density with a nuclear target
and that with a target nucleon:
\begin{equation}
\frac{dn/dy(p-A)}{dn/dy(p-p)}=A^{\beta(z)},
\end {equation} 
where y is the laboratory rapidity and z=y/ln(s).
  
The HEMAS p-Air model interaction assumes a scaling violation in the
central region and a small violation in the forward region ($x_{f}$$>$0.5).
The original HEMAS code included a naive muon transport code. This code
was later replaced with the more sophisticated PROPMU code\cite{propmu}.
Morever in 1995, HEMAS was interfaced with DPMJET, a dual parton model 
inspired code\cite{dpmjet}. The user has therefore the possibility of
changing the 
original HEMAS hadronic interaction model with DPMJET.
As far as the CPU time is concerned HEMAS is a fast code. Table 1 shows the
CPU time required for protons of different energies, while Table 2 shows the
comparison with other codes for a 200 TeV proton. 

\begin{table*} [hbt]
\begin{center}
\begin {tabular} {|l|l|}
\hline
$E_{p}(TeV)$        & CPU(HP-UX 9000)     \\
\hline
 20          &  0.01 sec/event             \\
\hline
200     &      0.17 sec/event         \\
\hline
200     &      0.93 sec/event\\
\hline
\end{tabular}
\caption{\it HEMAS CPU time for protons with different energies}
\label{tab:magne}
\end{center}
\end{table*}       
\begin{table*} [hbt]
\begin{center}
\begin {tabular} {|l|l|}
\hline
Code        & CPU     \\
\hline
 HEMAS        &  0.01 sec/event          \\
\hline
HEMAS-DPMJET  &      6.8 sec/event         \\
\hline
CORSIKA-QGS   &      3.4 sec/event\\
\hline
CORSIKA-SIBYLL&      2.9 sec/event\\
\hline
\end{tabular}
\caption{\it Comparison of the required CPU time for different
codes for 200 TeV protons}
\label{tab:magne}
\end{center}
\end{table*}       
An explanation of the faster performance of HEMAS, respect to other codes,
is in the treatment of the electromagnetic part of the shower.
Electromagnetic particles($e^{+}$,$e^{-}$,$\gamma$), coming from
$\pi_{0}$ decay, are computed using the standard NKG formula. Hadrons
falling below a given threshold are not transported in atmosphere and
their contribution to the electromagnetic size $N_{e}$, 
is computed according with the parametrization of 
pre-computed Monte Carlo runs\cite{gaibook}.
Of course the threshold is 
high enough ($E_{th}$$\simeq$500 GeV) 
to follow the hadrons until they can decay into an high energy muon, with some 
probability to survive deep underground.
Anyway, as far as the validity of this approximation is concerned,
it must be stressed that 
for primary cosmic rays with energy greater than $\simeq$10 TeV, the total
contribution of low energy hadrons to the electromagnetic size is 
$\simeq$10$\%$. 
\section{Comparison with experimental data}
The HEMAS code has been widely used to simulate the underground muons
detected at Gran Sasso. When dealing with underground muons, many 
experimental observables depend both on the 
cosmic ray chemical composition and on the 
features of the hadronic interaction model.
 To test the reliability of the Monte Carlo codes it's therefore 
important to study observables allowing a disentangle. The shape of the 
decoherence, i.e. the distribution of the distance between muon pairs, is
weakly dependent on C.R. composition. This distribution is therefore 
a nice test to check the reliability of a Monte Carlo code. 
The decoherence gets contribution from various sources in the shower
development:\\
- The primary cosmic ray cross section;\\
- the $p_{t}$ distribution of the muon parent hadrons;\\
- the multiple scattering of muons throught the rock.

Fig. 2 shows the average $p_{t}$ of the muon parent mesons as a function
of the average muon separation deep underground. The correlation between
$p_{t}$ and $<$D$>$ is evident.

The MACRO detector\cite{macro} is a powerfull experiment to study such 
distribution, 
taking advantage of an acceptance A$\simeq$10,000 $m^{2}$sr.
Recent results have been presented in\cite{icrc}: the decoherence function has
been studied with a statistical sample of 
$\simeq$ 350,000 real and 690,000 simulated muon pairs.
Fig.3 shows the comparison
between HEMAS expectation(MACRO composition model\cite{newpaper}) 
and the MACRO data, properly corrected for the
detector effects: the agreement is impressive.
The selection of high muon multiplicity events allows to study very high 
energy primary cosmic rays. Muons with multiplicity $N_{\mu}$$\geq$8, come from primary cosmic rays
with energy E$\geq$1000 TeV. The HEMAS expectation reproduces well
the experimental data of this subsample of events too( Fig. 4).
The two extreme composition models used are taken from\cite{auriemma}. 
 The comparison between data and Monte Carlo has been performed also
in different windows of rock depth and cos$\theta$. Fig. 5 shows the 
average distance between muon pairs in these windows: again HEMAS reproduces
quite well the experimental data.  

Summarizing, the MACRO data showed that, as far as the lateral distribution
of underground muons is concerned, the HEMAS capability in reproducing
the real data is impressive. 
Some doubts pointed out by the HEMAS authors of a possible $p_{t}$ excess 
in the code are not supported by the MACRO data \cite{gaisser}.

Neverthless, since the indirect measurements
aim to study the primary cosmic ray spectrum and composition, a
delicate sector of Monte Carlo simulation tools is the "absolute" muon flux.
It is of course an hard task to test experimentally the performance
of the Monte Carlo codes, since the muon flux deep underground 
is the convolution of the cosmic ray spectrum and composition with 
the hadronic interaction and the shower development features. Since the 
Cosmic Ray spectrum is unknown we cannot use the muon flux deep 
underground to test the Monte Carlo.

A step forward in this direction has been carried out by the
MACRO and EAS-TOP Collaborations, with the so called "anti-coincidences"
analysis\cite{spinetti}.
By selecting a muon events in MACRO pointing to a fiducial area well internal
to the EAS-TOP edges, it's possible to select two event samples:\\
a) if the number of fired detectors $N_{f}$ in EAS-TOP is $<$4, EAS-TOP
does not provide any trigger and the event is flagged as 'anti-coincidence'.
The correspondig C.R. energy ranges between 2 and few tens of TeV;\\
b) if 4$<$$N_{f}$$<$7, EAS-TOP provides a trigger and the events is 
flagged as 'low energy coincidences'.

In the energy range covered by 'anti-coincidences' and 'low energy
coincidences' direct measurements of cosmic ray spectrum and composition
are available. It is therefore possible to use these data as input to the
Monte Carlo simulation to test the hadronic interaction model,
by comparing the experimental data with the expectation. 
They used a single power low fits to the fluxes of H and He, as reported
by JACEE\cite{jacee}.
\begin{equation}
p:~~5.574\cdot10^{4}(E/GeV)^{-2.86}(m^{-2}s^{-1}sr^{-1}GeV^{-1})
\end{equation}
\begin{equation}
He: 9.15\cdot10^{3}(E/GeV)^{-2.86}(m^{-2}s^{-1}sr^{-1}GeV^{-1})
\end{equation}

I stress that this analysis cannot be performed with
MACRO alone, since low muon multiplicity events get contribution also 
from higher energy cosmic ray(E$\geq$100TeV), where the spectrum and
the chemical composition have not been measured with direct techniques.

Table 3 shows the results of the analysis and the comparison between the real
data and the Monte Carlo codes HEMAS and HEMAS-DPMJET.
\begin{table*} [t]
\begin{center}
\begin {tabular} {|l|l|l|}
\hline
  EAS-TOP fired &  $N_{f}$$<$4  & 4$\leq$$N_{f}$$<$7\\
 modules $N_{f}$&anti-coincidences& low energy coincidences \\
\hline
Exp. data        & 4239       &  376       \\
\hline
HEMAS            &   2729     &  324       \\
\hline
HEMAS-DPMJET     &  3502      &   314        \\
\hline
\end{tabular}
\caption{Comparison between the measured number of events with $N_{f}$
modules fired in EAS-TOP(triggered by a muon in MACRO) and the expectations
from two interaction models}.
\end{center}
\end{table*}       
Taking into account a 15-20$\%$ uncertainty in the JACEE data fits, the 
low energy coincidences are reproduced by both the Monte Carlo codes.
On the contrary HEMAS understimates the number of anti-coincidences
respect to the real data, while, within a 20$\%$ accuracy, HEMAS-DPMJET reproduces
the experimental data.

Sometime people is concerned by the fact that the HEMAS hadronic interaction
model reproduces the experimental data at high energy(E$\geq$100 TeV)
better than at 
lower energies,(1 TeV$\leq$E$\leq$100 TeV) being the latter closer to the 
energy range already explored by accelerator experiments. 
It must be stressed
that muons produced from the interaction of cosmic rays with energy
E$\simeq$ few TeV, come
from the decay of pions with $x_{f}$$\simeq$$E_{\pi}$/$E_{o}$$\simeq$1.
This is the so called 'forward region',
poorly studied in accelerator experiments, requiring therefore
an extrapolation in the Monte Carlo.
As it has been stressed in \cite{battisv}, the higher muon flux in DPMJET 
in this kinematical region, reflects an intrinsic feature of this 
code, originating from the LUND treatment 
of the fast valence
''diquark'' fragmentation in the projectile.
Fig.6 shows the average number of muons survived deep undeerground 
(h=3400 hg/c$m^{2}$) as a function of the proton energy for different
Monte Carlo codes. The main difference between these codes are infact found 
at low energy,
where each code has to extrapolate the collider results with  
some algorithms.
From this point of view, models based on the Dual Parton Model, can in
principle take advantage of the limited number of free parameters,
avoiding, at least in part, delicate extrapolations.

\section{Conclusions}
HEMAS is a fast Monte Carlo code for the simulation of high energy muons 
and electromagnetic components of the air shower. MACRO data confirm the
HEMAS capability in reproducing the lateral distribution of muons detected
deep underground. 

The 'low -energy coincidences' analysis performed 
by the EAS-TOP and MACRO collaborations pointed out a satisfactory 
agreement with HEMAS and HEMAS-DPMJET codes, within the primary C.R. 
spectrum uncertainty;
the 'anti-coincidences' analysis suggested a possible 
HEMAS muon deficit 
at threshold energies ( $E_{o}$$\simeq$few TeV). An improvement of the
agreement is found when using HEMAS interfaced with DPMJET.
\section{Acknowledgments}
I would like to thank my colleagues of the MACRO Collaboration and especially
the members of the muon working group for the fruitful discussions. A special
thank to G. Battistoni, C.Forti, J. Ranft and M.Sioli for their cooperation and
suggestions.

\newpage
\begin{center}
\mbox{\epsfig{file=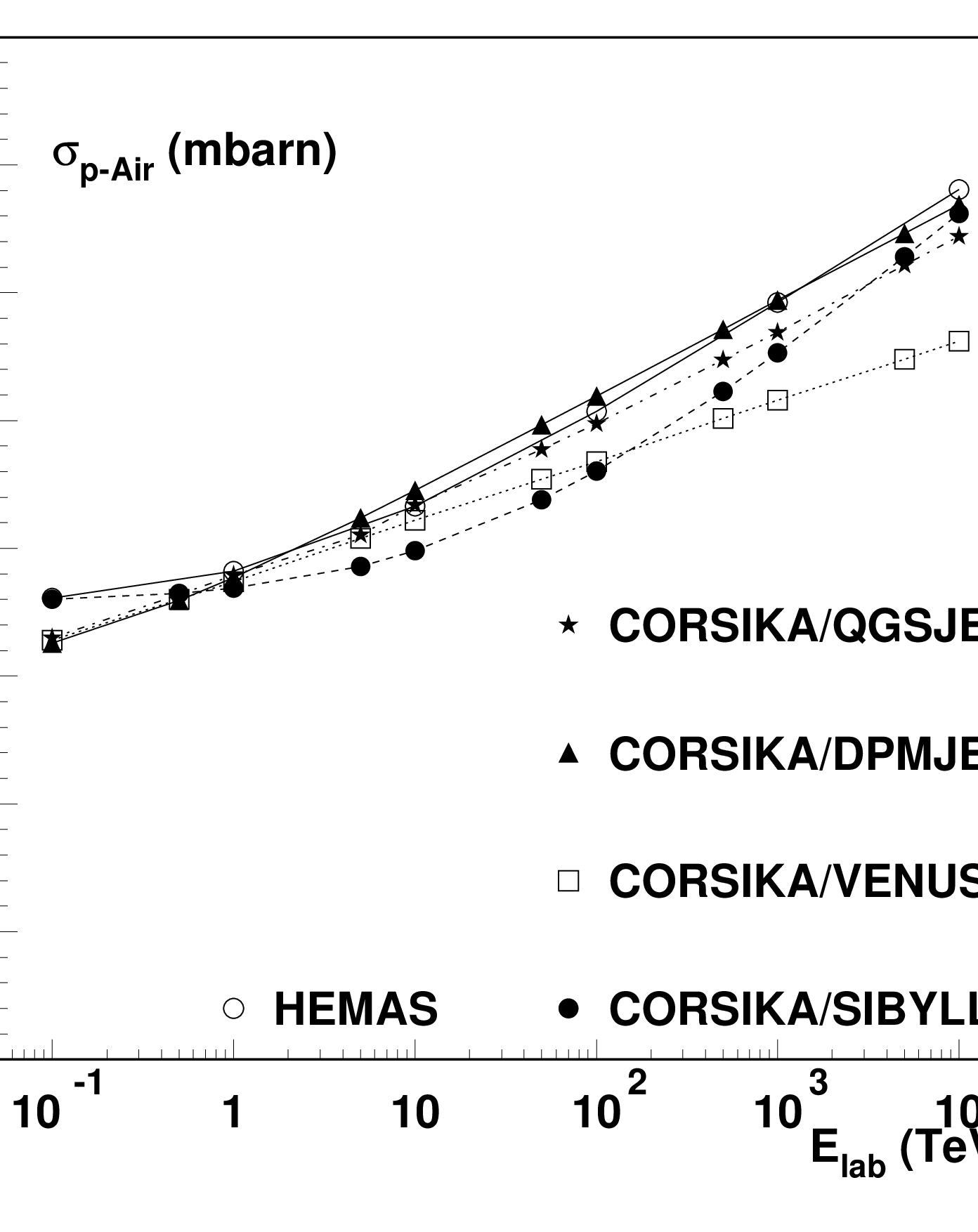,height=10cm,width=10cm}}
\vspace{2mm}
\noindent
\small
\end{center}
{\sf Figure~1:}~Comparison of the cross section p-Air used by different Monte
Carlo codes.
\begin{center}
\vskip 2. cm
\mbox{\epsfig{file=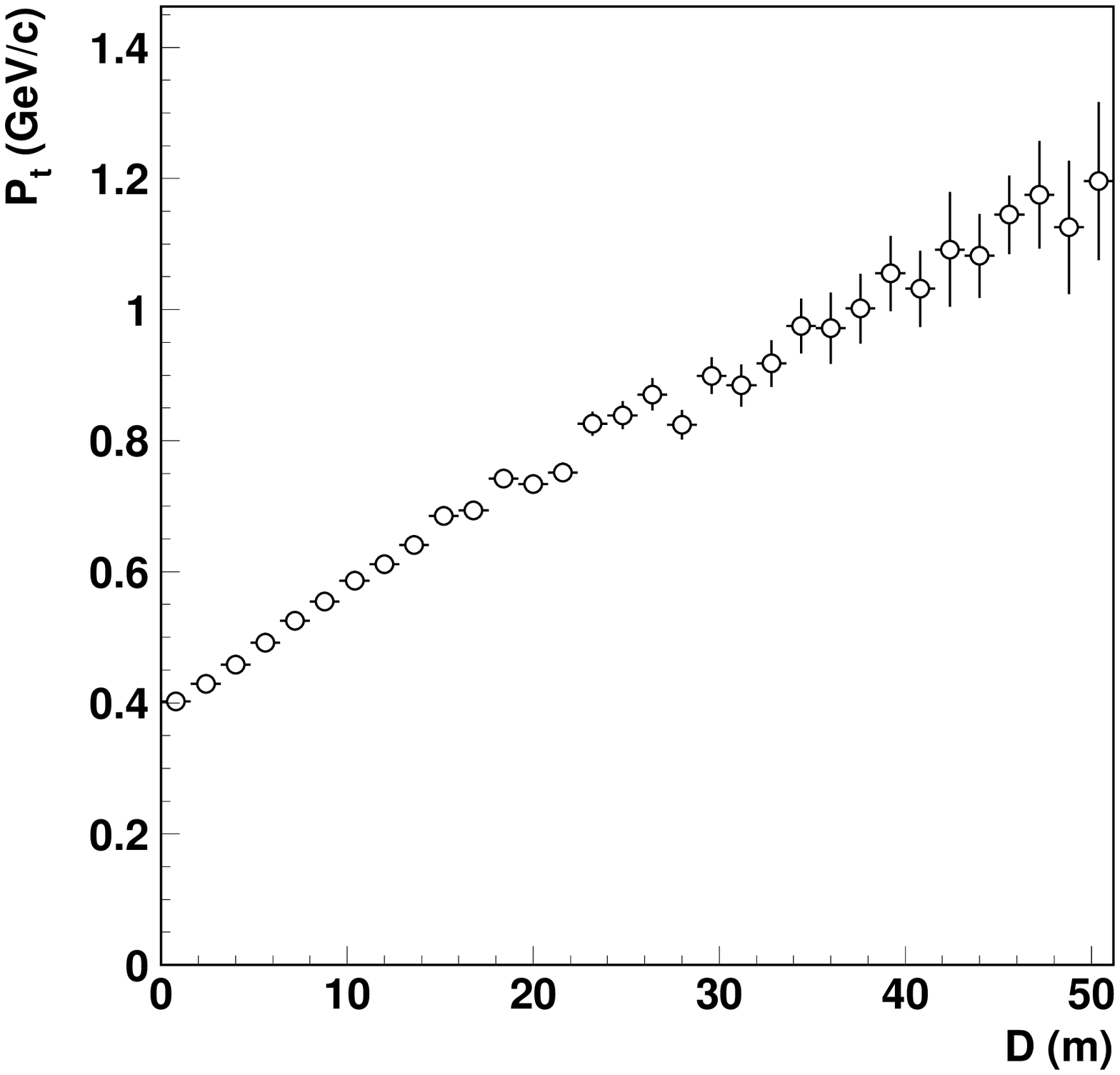,height=10cm,width=10cm}}
\vskip -1cm
\vspace{2mm}
\noindent
\small
\end{center}
{\sf Figure~2:}~Relation between the muon parent mesons average $P_{t}$
and the average muon pair distance deep underground.
\begin{center}
\vskip 2.5cm
\mbox{\epsfig{file=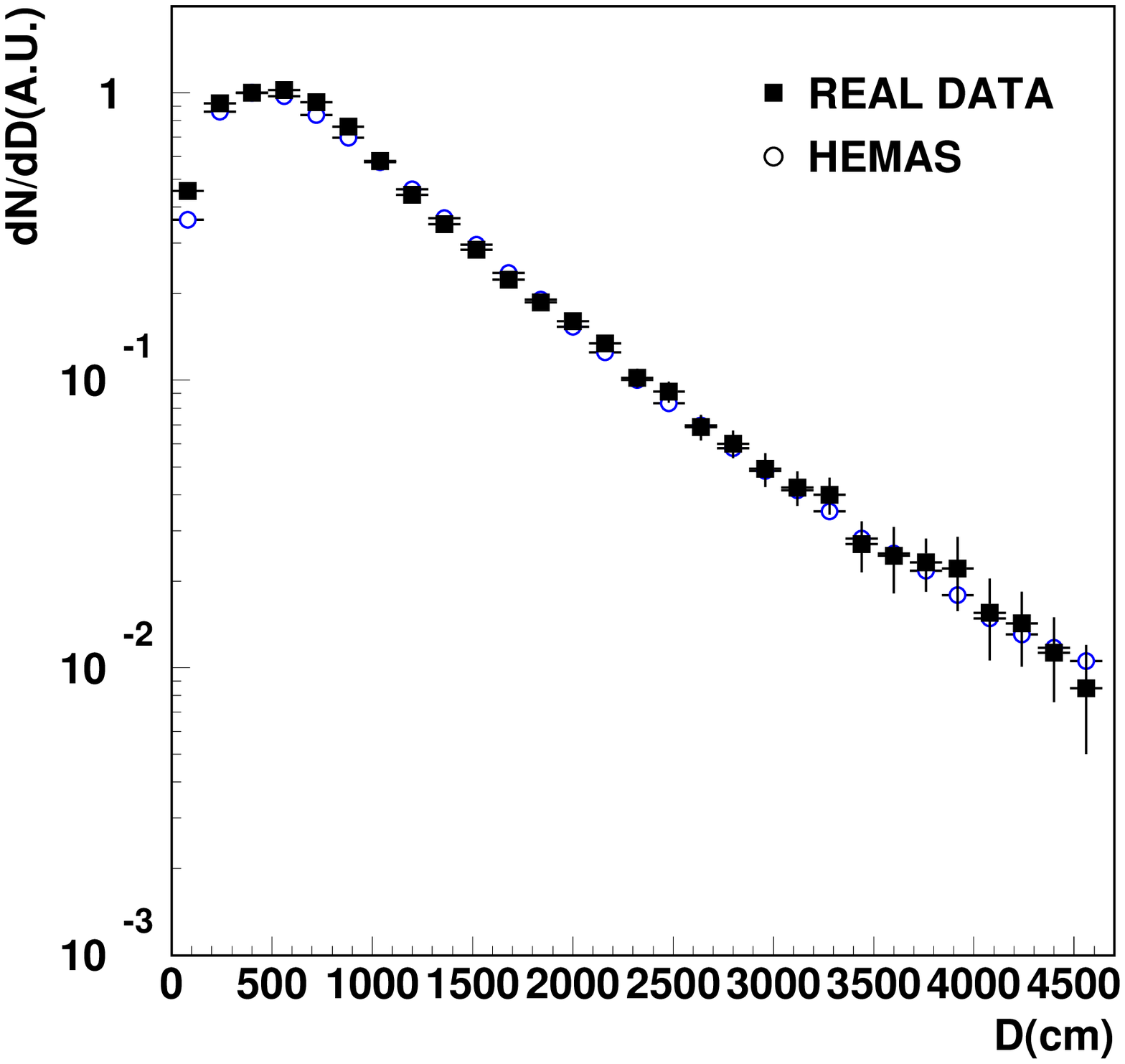,height=10cm,width=10cm}}
\vspace{2mm}
\vskip -1.cm
\noindent
\small
\end{center}
{\sf Figure~3:}~The decoherence function: comparison of the MACRO data 
with the HEMAS expectation.
\begin{center}
\vskip 0.5cm
\mbox{\epsfig{file=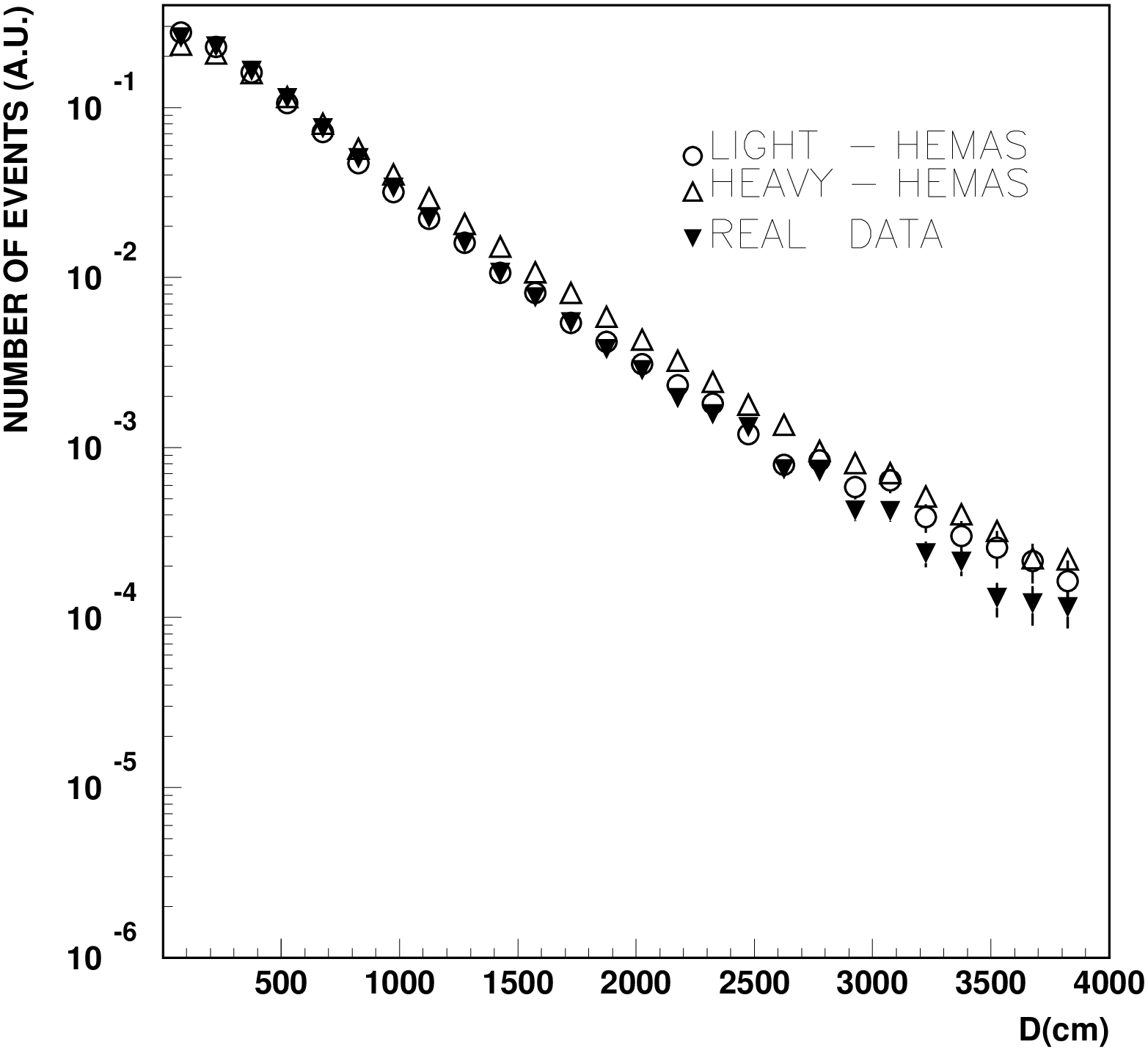,height=10cm,width=10cm}}
\vspace{2mm}
\noindent
\small
\end{center}
{\sf Figure~4:}~Comparison of the MACRO data and HEMAS expectation for 
events with muon multiplicity $N_{\mu}$$\geq$8.                                 
\begin{center}
\vskip 2.5cm
\mbox{\epsfig{file=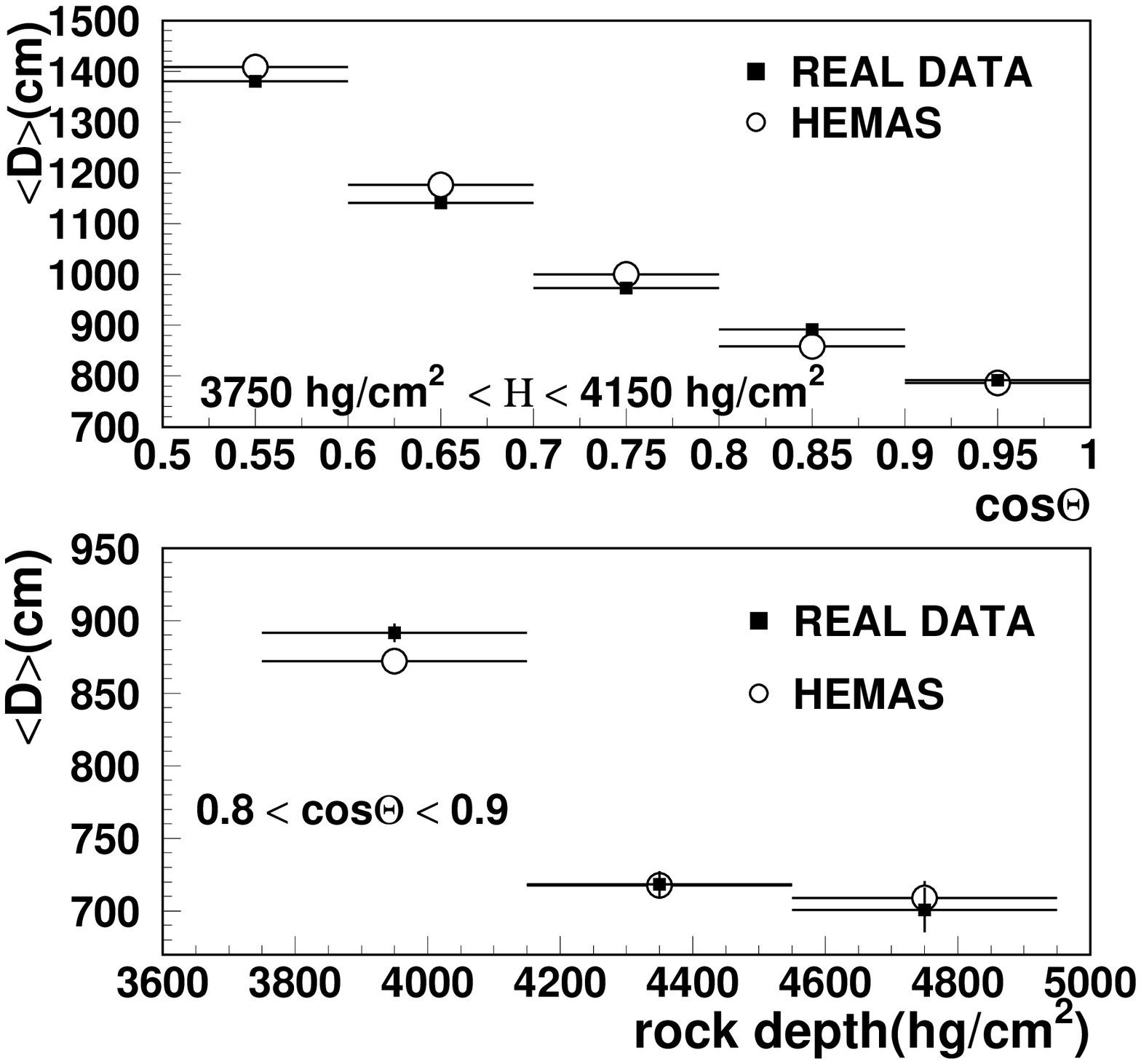,height=10cm,width=10cm}}
\vspace{2mm}
\noindent
\small
\end{center}
{\sf Figure~5:}~Comparison of the average separation between muon pairs
in different rock depth and cos$\theta$ windows.
\begin{center}
\vskip 0.5 cm
\mbox{\epsfig{file=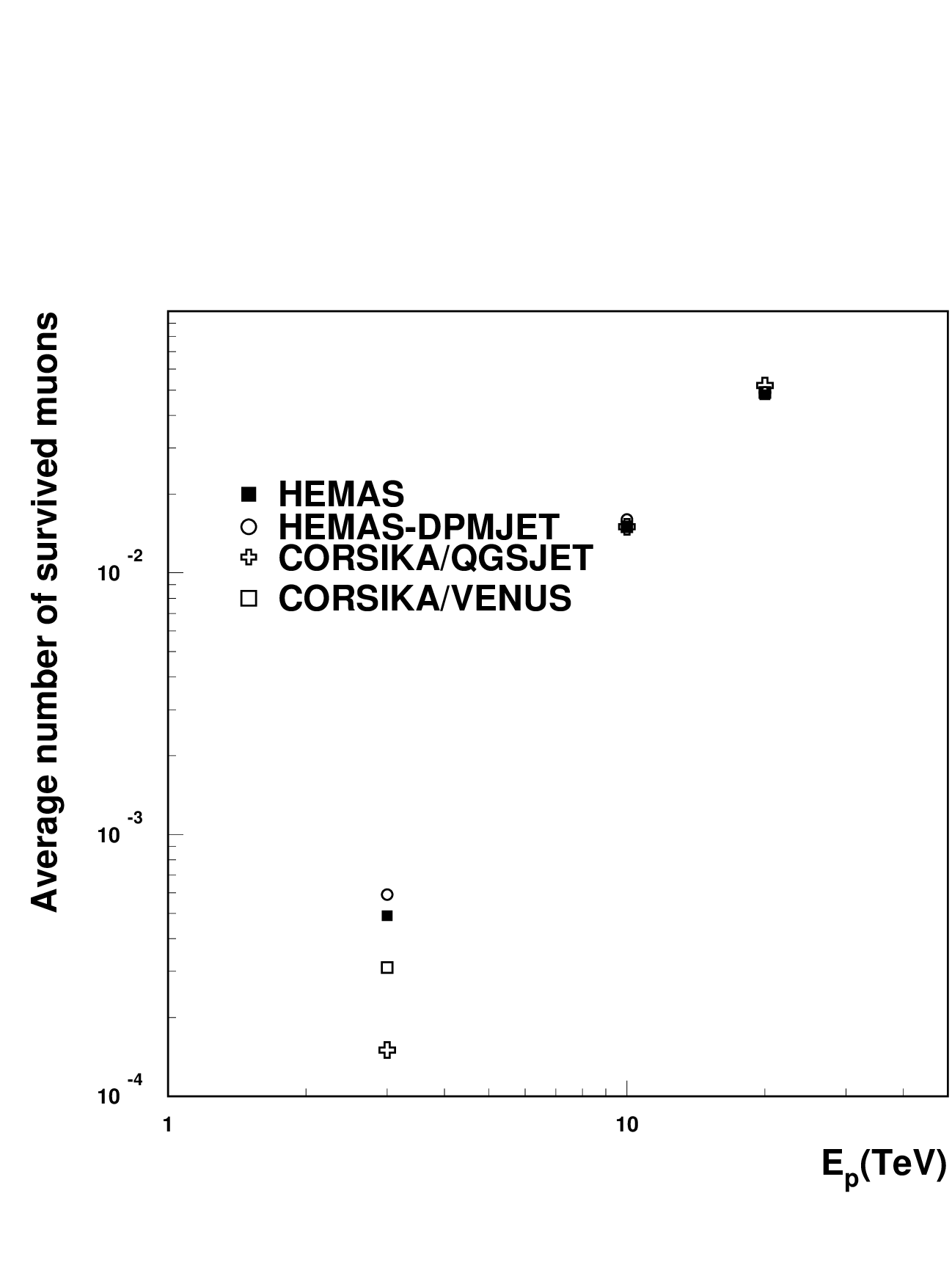,height=10cm,width=10cm}}
\vspace{2mm}
\noindent
\small
\end{center}
\vskip -1cm
{\sf Figure~6:}~Average number of muons survived underground (3400 hg/c$m^{2}$)
for different Monte Carlo codes as a function of proton energy. The same muon
transport (PROPMU) has been applied in all runs.
\normalsize   

\begin{thebibliography}{999}
%
\bibitem{hemas} C.~Forti {\it et al., Phys.~Rev.~D} {\bf 42} 3668 (1990).
\bibitem {cronin} J.W. Cronin {\it et al., Phys.~Rev.~D} {\bf 11} 3105 (1975).
\bibitem {propmu} P. Lipari and T. Stanev, Phys. Rev. {\bf D44} 3543 (1991).
\bibitem{gaibook} T. Gaisser, Cosmic Rays and particle physics, 
Cambridge Press, p.238
\bibitem{dpmjet} G. Battistoni et al.,Astroparticle Phys. {\bf 3} 157 (1995).
\bibitem{macro} The MACRO Coll.,{\it Nucl.~Instr.~Meth.~A} {\bf 324} 337 (1993).
\bibitem{icrc} The MACRO Coll., {\it Proc. 25rd ICRC,} Durban, {\bf 6} 357.
\bibitem{newpaper} The MACRO Coll. {\it Phys.~Rev.~D}{\bf 56} 1418 (1997).
\bibitem{auriemma} G.~Auriemma {\it et al., Proc.~22nd~ICRC,} Dublin,
    {\bf 2} 101 (1991).                                              
\bibitem {Palamara} The MACRO Coll.,{\it Proc. Taup95,}
Nucl. Phys. B(Proc. Suppl.), {\bf 48}, 444-446 (1996).
\bibitem{spinetti} The EAS-TOP and MACRO Collaborations, {\it Proc. 25rd ICRC,}
Durban,{\bf 6}, 85.                                                              
\bibitem {gaisser} T.K. Gaisser, Proc. Vulcano Workshop 1992, 40 433 (1993).
\bibitem{battisv} G. Battistoni,{\it Proc. of 10th ISVHECRI}, LNGS, in press.                                   
\bibitem{jacee}  The JACEE Collaboration 
{\it Proc. 23rd ICRC,} Calgary, {\bf 2} 25 (1993),
{\it Proc. 24rd ICRC,} Rome, {\bf 2} 728 (1995).

%
\normalsize
\end{thebibliography}
\end{document}